\documentclass[10pt]{article}
\usepackage{hyperref}
\usepackage{graphicx, amsmath, amsfonts, pseudo, chicago, setspace}

\def\st{:}                          

\newcommand{\ind}   [1]{1_{ \{ {#1} \} }}

\newcommand{\R}{\mathbb{R}}
\newcommand{\qed}{\nobreak \ifvmode \relax \else
      \ifdim\lastskip<1.5em \hskip-\lastskip
      \hskip1.5em plus0em minus0.5em \fi \nobreak
      $\Box$\fi}
\renewcommand{\algorithmiccomment}[1]{\hfill{\hbox to
8.9cm{\{ #1\}\hfill}}}

\newcommand{\minus}{\hbox to .15cm{--} }
\newcommand{\plus}{\hbox to .15cm{+} }

\newcommand{\PaperTitle}[1]{%
   \begin{center}%
      \begin{large}%
         \textbf {#1} \\%
      \end{large}%
   \end{center}%
}%

\newcommand{\AuthorList}[1]{%
   \begin{center}%
       \begin{large}%
           \textbf {#1} \\%
       \end{large}%
   \end{center}%
}%

\usepackage{geometry}%
\begin{document}
\doublespacing

\PaperTitle{Reduction Algorithm for the
 NPMLE for the Distribution Function\\ of Bivariate Interval Censored
 Data}

\AuthorList{Marloes H. Maathuis\footnote{Marloes H. Maathuis is a
Ph.D. student in the Department of Statistics, University of
Washington, Seattle, WA 98195 (email:
marloes@stat.washington.edu).}\\Department of Statistics,
University of Washington, Seattle, WA 98195 }

\abstract{We study computational aspects of the nonparametric
maximum likelihood estimator (NPMLE) for the distribution function
of bivariate interval censored data. The computation of the NPMLE
consists of two steps: a parameter reduction step and an
optimization step. In this paper we focus on the reduction step.
We introduce two new reduction algorithms: the Tree algorithm and
the HeightMap algorithm. The Tree algorithm is only mentioned
briefly. The HeightMap algorithm is discussed in detail and also
given in pseudo code. It is a very fast and simple algorithm of
time complexity $O(n^2)$. This is an order faster than the best
known algorithm thus far, the $O(n^3)$ algorithm of Bogaerts and
Lesaffre (2003). We compare our algorithms with the algorithms of
Gentleman and Vandal (2001), Song (2001) and Bogaerts and Lesaffre
(2003), using simulated data. We show that our algorithms, and
especially the HeightMap algorithm, are significantly faster.
Finally, we point out that the HeightMap algorithm can be easily
generalized to $d$-dimensional data with $d>2$. Such a
multivariate version of the HeightMap algorithm has time
complexity $O(n^d)$.
\\}

\noindent \textbf{Key words}: Computational Geometry; Maximal
Clique; Maximal Intersection; Parameter Reduction.\\

\begin{centering}\section{INTRODUCTION}\end{centering}

\noindent We consider the nonparametric maximum likelihood
estimator (NPMLE) for the distribution function of bivariate
interval censored data. Let $(X,Y)$ be the variables of interest
and let $F_0$ be their joint distribution function. Suppose that
there is a censoring mechanism, independent of $(X,Y)$, so that
$(X,Y)$ cannot be observed directly. Thus, instead of a
realization $(x,y)$, we observe a rectangular region $R \subset
\R^2$ that is known to contain $(x,y)$. We call $R$ an
\emph{observation rectangle}. Our data consists of $n$ i.i.d.\
observation rectangles $R_1,\dots,R_n$, and our goal is to compute
the NPMLE $\hat F_n$ of $F_0$.

Let $\mathcal F$ denote the class of all bivariate distribution
functions. Furthermore, for each $F\in \mathcal F$, let $P_F(R_i)$
denote the probability that the pair of random variables $(X,Y)$
is in observation rectangle $R_i$ when $(X,Y) \sim F$. Then,
omitting the part that does not depend on $F$, we can write the
log likelihood as
\begin{align} \label{eq:l_n}
  l_n(F)=\sum_{i=1}^n \log(P_F(R_i)),
\end{align}
and an NPMLE $\hat F_n \in \mathcal F$ is defined by
\begin{align*}
  l_n(\hat F_n) = \max_{F\in \mathcal{F}} \hspace{.1cm} l_n(F).
\end{align*}
As stated here, this is an infinite dimensional optimization
problem. However, the number of parameters can be reduced by
generalizing the reasoning of \citeN{Turnbull76} for univariate
censored data. By doing this (see e.g.
\citeN{BetenskyFinkelstein99}, \citeN{WongYu99},
\citeN{GentlemanVandal01}) one can easily derive that:
\begin{itemize}
  \item The NPMLE can only assign mass to a finite number of
    disjoint rectangles. We denote these rectangles by
    $A_1,\dots,A_m$ and call them \emph{maximal intersections},
    following \citeN{WongYu99}.
  \item The NPMLE is indifferent to the distribution of mass within
    the maximal intersections.
\end{itemize}
The second property implies that the NPMLE is non-unique, in the
sense that we cannot determine the distribution of mass within the
maximal intersections. \citeN{GentlemanVandal02} call this
\emph{representational non-uniqueness}. Hence, we can at best hope
to determine the amount of mass assigned to each maximal
intersection. Let $\alpha_j$ be the mass assigned to maximal
intersection $A_j$, and let $\alpha=(\alpha_1,\dots,\alpha_m)$.
Then $P_F(R_i)$ in equation (\ref{eq:l_n}) is simply the sum of
the probability masses of the maximal intersections that are
subsets of $R_i$:
\begin{align*}
  P_F(R_i) = P_{\alpha}(R_i) = \sum_{j=1}^m \alpha_j \ind{A_j \subset R_i}.
\end{align*}
We can then express the log likelihood in terms of $\alpha$:
\begin{align}\label{eq: log likelihood alpha}
  l_n(\alpha) & = \sum_{i=1}^n \log (P_{\alpha}(R_i)) =
  \sum_{i=1}^n  \log\Bigl( \sum_{j=1}^m \alpha_j \ind{A_j \subset
  R_i}\Bigr).
\end{align}
Let $\mathcal{K} = \left\{ \alpha \in \R^m: \alpha_j \ge 0,
j=1,\dots,m \right\}$ and $\mathcal{A} =\{\alpha\in \R^m \st
\mathbf{1}^T \alpha=1\}$, where $\mathbf{1}$ is the all-one vector
in $\R^m$. Then an NPMLE $\hat \alpha \in \mathcal K \cap \mathcal
A$ is defined by
\begin{align}\label{eq:red opt ind}
  l_n(\hat \alpha) = \max_{\alpha \in \mathcal K \cap \mathcal A}
  l_n(\alpha).
\end{align}
This is an $m$-dimensional convex constrained optimization problem
that does not need to have a unique solution in $\alpha$. This
forms a second source of non-uniqueness in the NPMLE.
\citeN{GentlemanVandal02} call this \emph{mixture non-uniqueness}.

Asymptotic properties of the NPMLE for univariate interval
censored data have been studied by \citeN{GroeneboomWellner92}. In
contrast to the consistency problems of the NPMLE for bivariate
right censored data, the NPMLE for bivariate interval censored
data has been shown to be consistent
(\citeN{VanderVaartWellner03:GC-preservationTheorems}). This
implies that the NPMLE can be used in practical applications to
estimate the distribution function of bivariate interval censored
data, for example to analyze data from AIDS clinical trials (see
e.g. \citeN{BetenskyFinkelstein99}).

From the discussion above, it follows that the computation of the
NPMLE consists of two steps. First, in the \emph{reduction step},
we need to find the maximal intersections $A_1,\dots,A_m$. This
reduces the dimensionality of the problem. Then, in the
\emph{optimization step}, we need to solve the optimization
problem defined in \eqref{eq:red opt ind}.\\

In this paper we focus on the reduction step. We distinguish
between two types of reduction algorithms that reflect a trade-off
between computation time and space:
\begin{itemize}
  \item type~1: The reduction algorithm computes the maximal
  intersections $A_1,\dots,A_m$.
  \item type~2: The reduction algorithm computes the \emph{clique matrix},
    an $m\times n$ matrix $C$ with elements $C_{ji}=\ind{A_j\subset
    R_i}$.
\end{itemize}
For $n$ observation rectangles, the number of maximal
intersections is $O(n^2)$. Hence, given the observation
rectangles, one can compute the clique matrix from the maximal
intersections and vice versa in $O(n^3)$ time.

We need $O(n^2)$ space to store the maximal intersections, while
we need $O(n^3)$ space to store the clique matrix. Thus, type~1
algorithms require an order of magnitude less space to store the
output. On the other hand, the choice of reduction algorithm
determines the amount of computational overhead in the
optimization step, where the values of the indicator functions
$\ind{A_j \subset R_i}$ are needed repeatedly. Namely, using a
type~1 algorithm requires repeated computation of these indicator
functions, while such computations are avoided with a type~2
algorithm. Thus, if we use a type~1 reduction algorithm, the
computational overhead in the optimization step is increased by a
constant factor.

Finally, it should be noted that the clique matrix provides useful
information about mixture uniqueness of the NPMLE. For example,
properties of the clique matrix are used to derive sufficient
conditions for mixture uniqueness by \citeN{GentlemanGeyer94} and
\citeN{GentlemanVandal02}. We can also use the clique matrix to
describe the equivalence class of solutions to \eqref{eq:red opt
ind}. Let $\hat \alpha$ be a solution, and consider $\left(C^T
\hat \alpha\right)_i = P_{\hat \alpha}(R_i)$, $i=1,\dots,n$. Since
the log likelihood \eqref{eq: log likelihood alpha} is strictly
concave in $P_{\alpha}(R_i)$, the vector $C^T \hat \alpha$ is
unique. Thus, the equivalence class of NPMLEs is exactly the set
$\{\alpha \in \mathcal K \cap \mathcal A \st C^T \alpha = C^T \hat
\alpha\}$, since all $\alpha$'s in this set yield the same
likelihood value.

We now give a brief overview of existing reduction algorithms.
\citeN{BetenskyFinkelstein99} provide a simple, but not very
efficient, type~1 algorithm. \citeN{GentlemanVandal01} introduce a
type~2 algorithm of time complexity $O(n^5)$. \citeN{Song01}
proposes a type~1 algorithm that is of comparable speed. The
algorithm with the best time complexity so far is the $O(n^3)$
type~1 algorithm of \citeN{BogaertsLesaffre03}. Finally,
\citeN{Lee83} gives an $O(n \log n)$ algorithm for a somewhat
different but related problem, namely that of finding the largest
number of rectangles having a non-empty intersection.

In this paper, we introduce two new reduction algorithms. The
algorithm we initially developed, the \emph{Tree} algorithm, is
only mentioned briefly. It is based on the algorithm of
\citeN{Lee83}, and is a fast but complex type~2 algorithm. Later,
we realized the reduction problem could be solved in a much
simpler way if one is only interested in finding the maximal
intersections. This resulted in the \emph{HeightMap} algorithm, a
very fast and simple type~1 algorithm of time complexity $O(n^2)$.
We discuss this algorithm in detail and also give it in pseudo
code. Finally, we compare the performance of our algorithms with
the algorithms of \citeN{GentlemanVandal01}, \citeN{Song01} and
\citeN{BogaertsLesaffre03}, using simulated data. We show that our
algorithms, and especially the HeightMap algorithm, are
significantly faster.

\begin{centering}\section{HEIGHT MAP ALGORITHM}\end{centering}

\noindent Recall that we want to find the maximal intersections
$A_1,\dots,A_m$ of a set of observation rectangles
$R_1,\dots,R_n$. There exist several equivalent definitions for
the concept of maximal intersection in the literature.
\citeN{WongYu99} use the following: $A_j\ne \emptyset$ is a
maximal intersection if and only if it is a finite intersection of
the $R_i$'s such that for each $i$ $A_j \cap R_i = \emptyset$ or
$A_j \cap R_i = A_j$. \citeN{GentlemanVandal02} use a graph
theoretic perspective: maximal intersections are the real
representations of maximal cliques in the intersection graph of
the observation rectangles.

\begin{figure}[t]
\begin{center}
    \includegraphics[width=.6\textwidth]{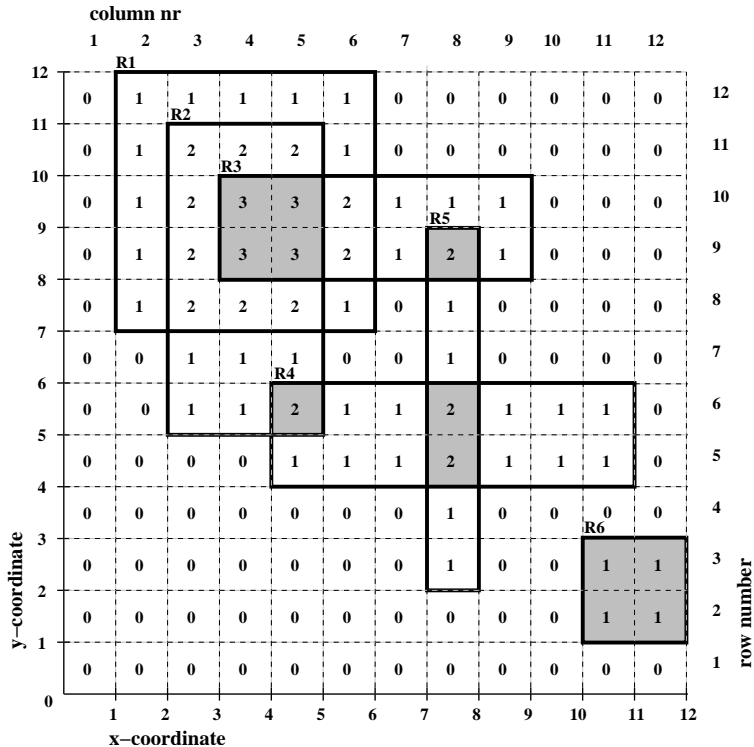}
    \caption{An example of six observation rectangles and their height map. The
    grey rectangles are the maximal intersections. Note that they
    correspond exactly to the local maxima of the height map.}
    \label{fig:canonrectheightmap}
\end{center}
\end{figure}

We view the maximal intersections in yet another way. We define a
\emph{height map} of the observation rectangles. This height map
is a function $h: \mathbb{R}^2 \rightarrow \mathbb{N}$, where
$h(x,y)$ is defined to be the number of observation rectangles
that contain the point $(x,y)$. The concept of the height map is
illustrated in Figure \ref{fig:canonrectheightmap}. It is easily
seen that the maximal intersections are exactly the local maxima
of the height map. This is true whenever there are no ties between
the observation rectangles, and this observation forms the basis
of our algorithm.\\

\subsection{Canonical rectangles}

\noindent We represent each rectangle $R_i$ as
$(x_{1,i},x_{2,i},y_{1,i},y_{2,i})$. The point $(x_{1,i},y_{1,i})$
is the lower left corner and $(x_{2,i},y_{2,i})$ is the upper
right corner of the rectangle. We call $(x_{1,i},x_{2,i}]$ the
$x$-interval, and $(y_{1,i},y_{2,i}]$ the $y$-interval of $R_i$.
Furthermore, we use boolean variables
$(c^x_{1,i},c^x_{2,i},c^y_{1,i},c^y_{2,i})$ to indicate whether an
endpoint is closed. As default we assume that left endpoints are
open and right endpoints are closed, so that
$(c^x_{1,i},c^x_{2,i},c^y_{1,i},c^y_{2,i}) = (0,1,0,1).$

We now transform the observation rectangles $R_1,\dots,R_n$ into
\emph{canonical rectangles} with the same intersection structure.
We call a set of $n$ rectangles canonical if all $x$-coordinates
are different and all $y$-coordinates are different, and if they
take on values in the set $\{1,2,\dots,2n\}$. An example of a set
of canonical rectangles is given in
Figure~\ref{fig:canonrectheightmap}.

\begin{figure}[t]
\begin{center}
    \includegraphics[width=.5\textwidth,trim=0 -15 0 5]{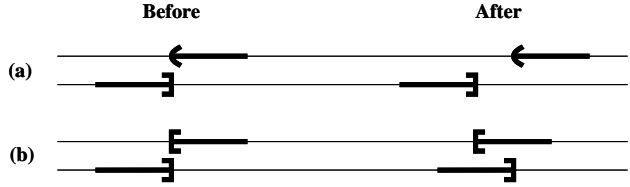}
    \caption{ Breaking ties during the transformation of observation
    rectangles into canonical rectangles.} \label{fig: examples canonical}
\end{center}
\end{figure}

We perform this transformation as follows. We consider the
$x$-coordinates and $y$-coordinates separately and replace them by
their order statistics. The only complication lies in the fact
that there might be ties in the data. Hence, we need to define how
to break ties. We explain the basic idea using the examples given
in Figure~\ref{fig: examples canonical}. In (a) we have an open
left endpoint $x_{1,i}$ and a closed right endpoint $x_{2,j}$ with
$x_{1,i}=x_{2,j}$ and $i\ne j$. Then the $x$-intervals of $R_i$
and $R_j$ have no overlap. Therefore, we sort the endpoints so
that the corresponding canonical intervals have no overlap, i.e.
we let $x_{2,j}$ be smaller. In (b) we have a closed left endpoint
$x_{1,i}$ and a closed right endpoint $x_{2,j}$ with
$x_{1,i}=x_{2,j}$ and $i\ne j$. Now the $x$-intervals of $R_i$ and
$R_j$ do have overlap. Therefore, we sort the endpoints so that
the corresponding canonical intervals overlap, i.e. we let
$x_{1,i}$ be smaller. In this way, we can consider all possible
combinations of endpoints. By listing the results in a table, we
found a compact way to code an algorithm for comparing endpoints.
It is given in pseudo code (Algorithm 1).

The reason for transforming the observation rectangles into
canonical rectangles is twofold. First, it forces us in the very
beginning to deal with ties and with the fact whether endpoints
are open or closed. As a consequence, we do not have to account
for ties and open or closed endpoints in the actual algorithm.
Second, it simplifies the reduction algorithm, since the column
and row numbers in the height map directly correspond to the $x$-
and $y$-coordinates of the canonical rectangles.

\subsection{Building the height map}

\noindent After transforming the rectangles, we build up the
height map. To this end, we use a sweeping technique commonly used
in the field of computational geometry (\citeNP{Lee83}). By using
this technique, we do not need to store the entire height map.
Instead, we only store one column at a time, in an array
$h_1,\dots,h_{2n}$. To build up the height map, we start with
$h_1,\dots,h_{2n}=0$. This is column 1 of the height map. We then
sweep through the plane, column by column, from left to right.
Every time we move to a new column, we either enter or leave one
observation rectangle. Thus, to compute the values of the height
map in the next column, we respectively increment or decrement the
values in the corresponding cells by 1. For example, when we move
from the first to the second column of the height map in Figure
\ref{fig:canonrectheightmap}, we enter rectangle $R_1$. $R_1$ has
$y$-interval $(7,12]$ which corresponds to rows 8 to 12 in the
height map. Hence, we increment $h_8,\dots,h_{12}$ by 1.\\

\subsection{Finding local maxima}

\noindent During the process of building up the height map, we can
find its local maxima, or equivalently, the maximal intersections.
We denote the maximal intersections in the same way as the
observation rectangles: $A_j=(x_{1,j},x_{2,j},y_{1,j},y_{2,j})$.
Suppose we apply the sweeping technique to the height map in
Figure \ref{fig:canonrectheightmap}, and suppose we are in column
5. We then are about to leave rectangle $R_2$. The $y$-interval of
$R_2$ is $(5,11]$, which corresponds to rows 6 to 11 in the height
map. Hence, the values of the height map will decrease by 1 in
rows 6 to 11, and will not change in the remaining rows. Since the
values of the height map are going to decrease, we may leave areas
of local maxima. Therefore, we need to look for local maxima in
rows 6 to 11 of column 5. We find two local maxima: the cell in
row 6, and the cells in rows 9 and 10. These local maxima in
column 5 correspond to local maxima in the height map, say $A_1$
and $A_2$ respectively. For $A_1$, we know that
$(y_{1,1},y_{2,1})=(5,6)$ and for $A_2$ we know that
$(y_{1,2},y_{2,2})=(8,10)$. Furthermore, from the fact that we
currently are in column 5, we know that $x_{2,1}=x_{2,2}=5$.
Finally, we obtain the values of $x_{1,1}$ and $x_{1,2}$ from the
left boundaries of the rectangles that were last entered. For the
cell in row 6 this is $R_4$ with left boundary 4. Hence, $A_1 =
(4,5,5,6)$. For the cells in rows 9 and 10, we last entered
rectangle $R_3$ with left boundary 3. Hence, $A_2=(3,5,8,10)$.
From this example we see that we need an additional array,
$e_1,\dots, e_{2n}$, where $e_k$ contains the index of the
rectangle that was last entered in row $k$ of the height map.

After finding the first local maxima we can continue the above
procedure. However, not every local maximum in the array $h$
corresponds to a local maximum in the complete height map. To
illustrate this problem, suppose that we are in column 6 of the
height map in Figure \ref{fig:canonrectheightmap}. We then are
about to leave rectangle $R_1$ with $y$-interval $(7,12]$.
Applying the above procedure, we look for local maxima in rows 8
to 12 of column 6, and we find a maximum in rows 9 and 10.
However, this does not correspond to a local maximum in the height
map. It merely is a remainder from the maximal intersection $A_2$
that we found earlier. Namely, the local maximum in column 6 is
formed by the set $\{R_1,R_3\}$ which is a subset of the set
$\{R_1,R_2,R_3\}$ that forms $A_2$. We can prevent the output of
such pseudo local maxima as follows. After we output a maximal
intersection $A_j$, we set $e_k := 0$ for one of the rows in
$A_j$. Then, a local maximum in the array $h$ corresponds to a
maximal intersection if and only if $e_k >0$ for all of its cells.
In the example in Figure \ref{fig:canonrectheightmap}, this means
that after we output $A_1$ and $A_2$ we need to set $e_k := 0$ for
one of their rows. $A_1$ only consists of row 6, and therefore we
set $e_6 := 0$. $A_2$ consists of rows 9 and 10, and we choose to
set $e_9 :=0$. Then, when we find the local maximum in rows 9 and
10 of column 6, we know it does not correspond to a maximal
intersection since $e_9 = 0$.

Summarizing, we sweep through the plane from left to right, column
by column. At each step in the sweeping process we either enter or
leave a canonical rectangle. When we enter a rectangle $R_i$ with
$y$ interval $(y_{1,i},y_{2,i}]$, we increment $h_k$ by 1 and set
$e_k:=i$ for $k= y_{1,i}+1,\dots,y_{2,i}$. When we leave a
rectangle $R_i$, we first look for local maxima in $h_k$ for
$k=y_{1,i}+1,\dots,y_{2,i}$. For each local maximum that we find
in $h$, we check whether $e_k > 0$ for all of its cells. If this
is the case, we output the corresponding maximal intersection and
set $e_k := 0$ for one of the cells in the local maximum. Finally,
we decrement $h_k$ by 1 for $k= y_{1,i}+1,\dots,y_{2,i}$. The
complete algorithm is given in pseudo code (Algorithm 2). An
R-package of the algorithm is available at
http://www.stat.washington.edu/marloes.

\subsection{Time and space complexity}

\noindent We can easily determine the time and space complexity of
the algorithm. In order to transform a set of rectangles into
canonical rectangles, we need to sort the endpoints of their
$x$-intervals and $y$-intervals. This takes $O(n \log n)$ time and
$O(n)$ space. At each step in the sweeping process, we need to
update at most $2n$ cells of the arrays $h$ and $e$. Furthermore,
we may need to find local maxima in at most $2n$ cells, and we may
need to check whether $e_k >0$ for at most $2n$ cells. Thus, the
time complexity of one sweeping step is $O(n)$. Combining this
with the fact that the number of sweeping steps is $O(n)$ gives a
total time complexity of $O(n^2)$. With respect to the space
complexity, we need to store the arrays $h$ and $e$. Hence, the
space complexity for computing the maximal intersections is
$O(n)$. However, storing the maximal intersections takes $O(n^2)$
space.

\begin{centering}\section{EVALUATION OF THE ALGORITHMS}\end{centering}

We compared our algorithms with the algorithms of
\citeN{GentlemanVandal01}, \citeN{Song01}, and
\citeN{BogaertsLesaffre03}, using simulated data. We generated
bivariate current status data according to a very simple
exponential model:
\begin{align}\label{model:expbcs}
 X,Y,U,V \sim \exp(1),
\end{align}
where $X$ and $Y$ are the variables of interest, $U$ is the
observation time for $X$, $V$ is the observation time for $Y$, and
$X$, $Y$, $U$ and $V$ are mutually independent. Thus, the
observation rectangles were generated as follows:
\begin{align*}
  \begin{array}{llll}
   X_i \le U_i, & Y_i \le V_i & \Rightarrow & R_i=(0,U_i,0,V_i) \\
   X_i \le U_i, & Y_i > V_i & \Rightarrow & R_i =
     (0,U_i,V_i,\infty) \\
   X_i > U_i, & Y_i \le V_i & \Rightarrow & R_i =
     (U_i,\infty,0,V_i) \\
   X_i > U_i,& Y_i > V_i & \Rightarrow & R_i =
   (U_i,\infty,V_i,\infty)
  \end{array}
\end{align*}
We used sample sizes $50$, $100$, $250$, $500$, $1,\!000$,
$2,\!500$, $5,\!000$ and $10,\!000$. For each sample size, we ran
50 simulations on a Pentium IV 2.4GHz computer with 512 MB of RAM
and we recorded the user times of the algorithms. For each
algorithm, we omitted sample sizes that took over $1,\!000$
seconds to run. All algorithms were implemented in C.

The results of the simulation are shown in Table~\ref{table:
results simulation}. We see that the Tree algorithm, and
especially the HeightMap algorithm are significantly faster than
the other algorithms. The HeightMap algorithm runs sample sizes of
$10,\!000$ in less than two seconds.

\begin{table}
\centering
\begin{small}
 \begin{tabular}{|r | r@{.}l r@{.}l | r@{.}l r@{.}l | r@{.}l r@{.}l | r@{.}l r@{.}l | r@{.}l
                 r@{.}l|}
   \hline
 & \multicolumn{4}{c|}{Gentleman\&Vandal} &
  \multicolumn{4}{c|}{Song} &
  \multicolumn{4}{c|}{Bogaerts\&Lesaffre} &
  \multicolumn{4}{c|}{Tree} &
  \multicolumn{4}{c|}{HeightMap} \\
   $n$ & \multicolumn{2}{c}{mean} & \multicolumn{2}{c|}{sd}
        & \multicolumn{2}{c}{mean} & \multicolumn{2}{c|}{sd} &
        \multicolumn{2}{c}{mean} & \multicolumn{2}{c|}{sd}  &
        \multicolumn{2}{c}{mean} & \multicolumn{2}{c|}{sd}  &
        \multicolumn{2}{c}{mean} & \multicolumn{2}{c|}{sd}  \\
 \hline
 50   & 0&0004 & 0&0028 & 0&029 & 0&011 & 0&0010 & 0&0042 & 0&0012 & 0&0044 & 0&0006 & 0&0031 \\

 100  & 0&001 & 0&0036 & 0&52 & 0&14 & 0&0052 & 0&0079 & 0&0036 &
 0&0072 & 0&0008 & 0&0040 \\

 250  & 0&061 & 0&015 & 26&0 & 47&0 & 0&083 & 0&014 & 0&016 & 0&0053
 & 0&0018 & 0&0056 \\

 500  & 1&3 & 0&48 & 540&0 & 100&0 & 0&91 & 0&11 & 0&058 & 0&0087
 & 0&0060 & 0&0083 \\

 1,000 & 46&0 & 63&0 & \multicolumn{2}{c}{NA} &
 \multicolumn{2}{c|}{NA} & 13&0 & 1&0 & 0&29 & 0&032 & 0&019 & 0&0082 \\

 2,500 & \multicolumn{2}{c}{NA} & \multicolumn{2}{c|}{NA}  & \multicolumn{2}{c}{NA} & \multicolumn{2}{c|}{NA}
       & 470&0 & 30&0 & 3&1 &
       0&10 & 0&10 & 0&011 \\

 5,000 & \multicolumn{2}{c}{NA} & \multicolumn{2}{c|}{NA}  & \multicolumn{2}{c}{NA} & \multicolumn{2}{c|}{NA}
       & \multicolumn{2}{c}{NA} & \multicolumn{2}{c|}{NA}  & 25&0 &
       0&37  & 0&38 & 0&014 \\

 10,000 & \multicolumn{2}{c}{NA} & \multicolumn{2}{c|}{NA}  & \multicolumn{2}{c}{NA} & \multicolumn{2}{c|}{NA}
       & \multicolumn{2}{c}{NA} & \multicolumn{2}{c|}{NA}  & 180&0 &
       2&7  & 1&4 & 0&029 \\

 \hline
 \end{tabular}
 \caption{Mean and standard deviation of the user time in seconds,
 over 50 simulations per sample size from model
 \eqref{model:expbcs}. Cells with NA indicate
 that simulations took over 1,000 seconds to run and were
 therefore omitted.}
 \label{table: results simulation}
 \end{small}
\end{table}

\begin{figure}[t]
\begin{center}
    \includegraphics[width=.6\textwidth,trim=0 0 0 15]{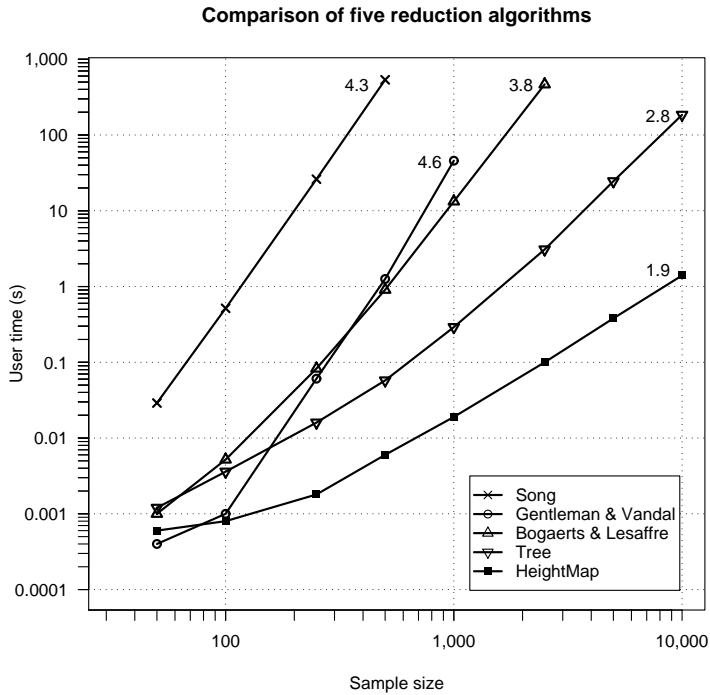}
    \caption{ Log-log plot of the mean user time in seconds versus the sample
    size, over 50 simulations per sample size from model~\eqref{model:expbcs}.
    For each algorithm, the estimated slope of its graph is given. These slopes
    can be used as empirical estimates of the time complexity of the algorithms. }
    \label{fig:reduc algorithms}
\end{center}
\end{figure}

To get an empirical idea of the time complexity of the algorithms,
Figure \ref{fig:reduc algorithms} shows a log-log plot of the mean
user time versus the sample size. We fitted least squares lines
through the last 4 points of each algorithm. The slopes of these
lines can be used as empirical estimates of the time complexity of
the algorithms. We see that the estimated slope of the HeightMap
algorithm is 1.9, which agrees with the theoretical time
complexity of $O(n^2)$ that we derived earlier. Furthermore, we
see that the HeightMap algorithm is about an order faster than the
Tree algorithm, which is about an order faster than the algorithm
of Bogaerts and Lesaffre. Finally, note that the empirical time
complexity of the algorithm of Bogaerts and Lesaffre is greater
than the theoretical complexity of $O(n^3)$ that they derived.\\

\begin{centering}\section{MULTIVARIATE HEIGHTMAP ALGORITHM}\end{centering}

The height map algorithm can be easily generalized to higher
dimensional data. For example, for 3-dimensional interval censored
data the observation sets $R_i$ take the form of 3-dimensional
blocks $(x_{1,i},x_{2,i},y_{1,i},y_{2,i},z_{1,i},z_{2,i})$. In
this situation the height map is a function $h: \R^3 \to \mathbb
N$, where $h(x,y,z)$ is the number of observation sets that
contain the point $(x,y,z)$. The maximal intersections again
correspond to local maxima of the height map. By first
transforming the observation sets into canonical sets, this
implies that we need to find the local maxima of a $2n \times 2n
\times 2n$ matrix. We can do this by sweeping through the matrix,
slice by slice, say along the $z$-coordinate. We only store one
slice of the height map at a time, so that $h$ and $e$ are now $2n
\times 2n$ matrices. At each step in the sweeping process, we
either enter or leave an observation set $R_i$. When we enter an
observation set, we update the corresponding values of $h$ and
$e$, i.e. we set $h_{k,l} := h_{k,l} + 1$ and $e_{k,l} := i$ for
all $k=x_{1,i}+1,\dots,x_{2,i}$ and $l=y_{1,i}+1,\dots,y_{2,i}$.
When we leave an observation set, we look for local maxima in the
cells of the rectangle $(x_{1,i},x_{2,i},y_{1,i},y_{2,i})$, using
the height map algorithm for 2-dimensional data. For each local
maximum that we find, we check whether $e_{k,l}>0$ for all of its
cells. If this is the case, we output the corresponding maximal
intersection and set $e_{k,l} := 0$ for one of the cells in the
local maximum. Finally, we decrement $h_{k,l}$ by 1 for $k=
x_{1,i}+1,\dots,x_{2,i}$ and $l = y_{1,i}+1,\dots,y_{2,i}$.

For $d$-dimensional data, the time complexity of a sweeping step
is $O(n^{d-1})$. Since the number of sweeping steps is $O(n)$,
this gives a total time complexity of $O(n^d)$. With respect to
the space complexity, we need to store the matrices $h$ and $e$.
Hence, the space complexity to compute the maximal intersections
is $O(n^{d-1})$. However, storing the maximal intersections takes
$O(n^d)$ space.\\

\begin{centering}\section*{ACKNOWLEDGEMENTS}\end{centering}

\noindent This research was partly supported by NSF grant
DMS-0203320. The author would like to thank Kris Bogaerts,
Shuguang Song, and Alain Vandal for providing the code of their
algorithms, and a referee for suggesting to consider generalizing
the HeightMap algorithm to $d$-dimensional data. Finally, the
author would like to thank Piet Groeneboom, Steven Schimmel and
Jon Wellner for their contributions, support and encouragement.\\

\begin{centering}\section*{APPENDIX: PSEUDO CODE}\end{centering}
\begin{pseudocode}%
  {CompareEndpoints}%
  {$A$,$B$}%
  {Two endpoint descriptors $A=(x_{k,i},c^x_{k,i})$ and $B=(x_{l,j},c^x_{l,j})$}%
  {A boolean value indicating $A < B$}
  \STATE{$c_A := (c^x_{k,i}=1)$} \COMMENT{boolean indicating
  $A$ is a closed endpoint}
  \STATE{$c_B:= (c^x_{l,j}=1)$} \COMMENT{boolean indicating $B$
  is a closed endpoint}
  \STATE{$r_A := (k=2)$} \COMMENT{boolean indicating $A$ is a
  right endpoint}
  \STATE{$r_B:= (l=2)$} \COMMENT{boolean indicating $B$ is a right
  endpoint}
  \IF[if the endpoints have different coordinates]{($x_{k,i} \ne x_{l,j}$)}
    \STATE{\RETURN ($x_{k,i} < x_{l,j}$)}
      \COMMENT{\dots then let their coordinates determine their order}
  \ENDIF
  \IF[if the endpoints are identical]{($r_A=r_B$ \textbf{and} $c_A =c_B$)}
    \STATE{\RETURN ($i < j$)}
      \COMMENT{\dots then let their index determine their order}
  \ENDIF
  \IF[if the endpoints are opposites]{($r_A \ne r_B$ \textbf{and} $c_A  \ne c_B$)}
    \STATE{\RETURN ($r_A$)}
      \COMMENT{\dots then $A<B$ when $A$ is a right endpoint}
  \ENDIF
  \STATE{ \RETURN $(r_A \ne c_A)$ }
    \COMMENT{otherwise $A < B$ when $A$ is closed left or open right}
\end{pseudocode}

\begin{pseudocode}{HeightMapAlgorithm2D}{$R_1,\dots,R_n$}{A
set of $n$ 2-dimensional observation rectangles $R_1,\dots,R_n$}
{The corresponding maximal intersections $A_1,\dots,A_m$}

\STATE{Transform observation rectangles into canonical rectangles
$(x_{1,i},x_{2,i},y_{1,i},y_{2,i})$, using CompareEndpoints}

\STATE{Sort $x_{1,i},x_{2,i}$, $i=1,\dots,n$ in ascending order
and store their indices $i$ in the list $r_{1},\dots,r_{2n}$}

\STATE{$m:=0$} \COMMENT{counts number of maximal intersections}

\STATE{$h_{1},\dots,h_{2n}:=0$} \COMMENT{column of height map}

\STATE{$e_{1},\dots,e_{2n}:=0$} \COMMENT{index of last entered
rectangle; 0 blocks output}

\FOR[sweep through height map from column $1$ to $2n$]{$j=1$ to
$2n$}

    \IF [we enter a rectangle]{(${r_j}$ is a left boundary)}

        \FOR [update
        $h_k$ and $e_k$ for $k=y_{1,r_j}+1, \dots, y_{2,r_j}$] {$k=y_{1,r_j}+1$ \TO $y_{2,r_j}$}

            \STATE {$h_k:=h_k+1$; \quad $e_k:=r_j$}

        \ENDFOR

    \ELSE[we leave a rectangle]

        \STATE{$b:=y_{1,r_j}$} \COMMENT{bottom coordinate of local maximum; 0 blocks output}

        \FOR [look
        for local maxima in rows $y_{1,r_j} +1, \dots,
        y_{2,r_j}-1$]{$k=y_{1,r_j} +1$ \TO $y_{2,r_j}-1$}

            \IF[there is a local maximum in $h$] {($h_{k+1} < h_k$  \textbf{and} $b>0$)}

                \IF[the local maximum in $h$ is a maximal intersection] {$(e_{b+1},\dots,e_k > 0)$}

                    \STATE {$m:=m+1; \quad
                    A_m:=(x_{1,e_k},j,b,k)$} \COMMENT{output the maximal intersection}
                    \STATE { $e_{b+1} := 0$} \COMMENT{set $e$ to zero for row $b+1$ in $A_m$}

                \ENDIF

                \STATE {$b:=0$}

            \ENDIF

            \IF {($h_{k+1} > h_k$)}

                \STATE {$b:=k$}

            \ENDIF

        \ENDFOR

        \STATE{ $k:= y_{2,r_j}$}\COMMENT{look for local maximum in row $y_{2,r_j}$}

        \IF[there is a local maximum in $h$]{($b>0$)}

            \IF[the local maximum in $h$ is a maximal intersection] {$(e_{b+1},\dots,e_k>0)$}

               \STATE {$m:=m+1; \quad
                      A_m:=(x_{1,e_k},j,b,k)$} \COMMENT{output the
                      maximal intersection}
               \STATE{$e_{b+1} := 0$}\COMMENT{set $e$ to zero for row $b+1$ in $A_m$}
            \ENDIF

        \ENDIF

        \FOR [update $h_k$ for $k=y_{1,r_j}+1,\dots,y_{2,r_j}$] {$k=y_{1,r_j}+1$ \TO $y_{2,r_j}$}

            \STATE {$h_k:=h_k-1$}

        \ENDFOR

    \ENDIF

\ENDFOR

\STATE {Transform the canonical maximal intersections
$A_1,\dots,A_m$ back to the original coordinates}

\STATE{\RETURN $A_1,\dots,A_m$}
\end{pseudocode}

\end{document}